# ODIN-Based CPU-GPU Architecture with Replay-Driven Simulation and Emulation

Nij Dorairaj[1], Debabrata Chatterjee[1], Hong Wang[1], Hong Jiang[1], Alankar Saxena[1], Altug Koker[1], Thiam Ern Lim[1], Cathrane Teoh[1], Chuan Yin Loo[1], Bishara Shomar[2], Anthony Lester[3]

[1]Intel Corporation, [2]Intel Corporation/Nvidia Corp, [3]Synopsys Inc

**Abstract**

Integration of CPU and GPU technology is a key enabler for modern AI and graphics workloads, combining control-oriented processing with massive parallel compute capability. As systems evolve toward chiplet-based architectures, pre-silicon validation of tightly coupled CPU–GPU subsystems presents significant challenges due to complex validation framework setup, large design scale, high concurrency, non-deterministic execution, and intricate protocol interactions at chiplet boundaries—often resulting in long integration cycles. This paper presents a replay-driven validation methodology developed during the integration of a CPU subsystem, multiple Xe GPU cores, and a configurable Network-on-Chip (NoC) within a foundational SoC building block targeting the ODIN integrated chiplet architecture. By leveraging deterministic waveform capture and replay across both simulation and emulation using a single design database, complex GPU workloads and protocol sequences can be reproduced reliably at the system level. This approach significantly accelerates debug, improves integration confidence, and enables end-to-end system boot and workload execution within a sin-gle quarter, demonstrating the effectiveness of replay-based validation as a scalable methodology for chiplet-based systems.

## 1 Introduction

CPU–GPU integration has become foundational for System-on-Chip (SoC) designs targeting AI, media, and high-performance computing. While CPUs excel at control and latency-sensitive tasks, GPUs deliver high throughput through massive parallelism. Integrating these subsystems introduces validation challenges due to execution infrastructure requirements, multiple levels of memory hierarchies, complex interface protocols, large address maps, and non-deterministic behav-ior. In addition, traditional directed testing is insufficient to uncover integration-level issues that only manifest under realistic workloads. Replay-driven valida-tion bridges the gap between simulation visibility and emulation performance.

## 2 Odin SOC Overview and Chiplet Context

Modern high-performance SoC designs increasingly adopt a chiplet-based integra-tion model to improve scalability, modularity, and design reuse. In this approach, complex compute and memory subsystems are composed as semi-independent chiplets and integrated using well-defined interconnect and protocol boundaries. This work is based on an ODIN integrated chiplet, which brings together CPU, GPU, memory controllers, and system interconnect into a single composable unit. The ODIN integrated chiplet combines a Xeon CPU subsystem, a GPU compute complex, and a Network-on-Chip (NoC) fabric, along with high-bandwidth and standard memory interfaces. As shown in Figure 1, the CPU and GPU subsystems connect to the NoC fabric, which serves as the primary communication backbone for coherent and non-coherent traffic. The NoC also interfaces with external memory controllers, PCIe, and other SoC IPs, enabling system-level integration while preserving modularity at the chiplet boundary.

A key aspect of this architecture is the separation of compute chiplets from memory technologies. High-bandwidth memory (HBM) is accessed through a dedicated HBM controller within the chiplet, while traditional DDR memory is accessed via a DDR controller connected to the NoC. This separation allows the chiplet to support heterogeneous memory systems and enables flexible deployment across different platform configurations. By adopting a chiplet-based architecture, the ODIN design enables independent development and validation of CPU and GPU subsystems, while relying on the NoC fabric to provide scalable connectivity, ordering, and protocol translation. This modularity, while beneficial for integration and reuse, also introduces validation challenges—particularly at the interfaces between chiplets—which motivate the replay-driven validation methodology described in the following sections.

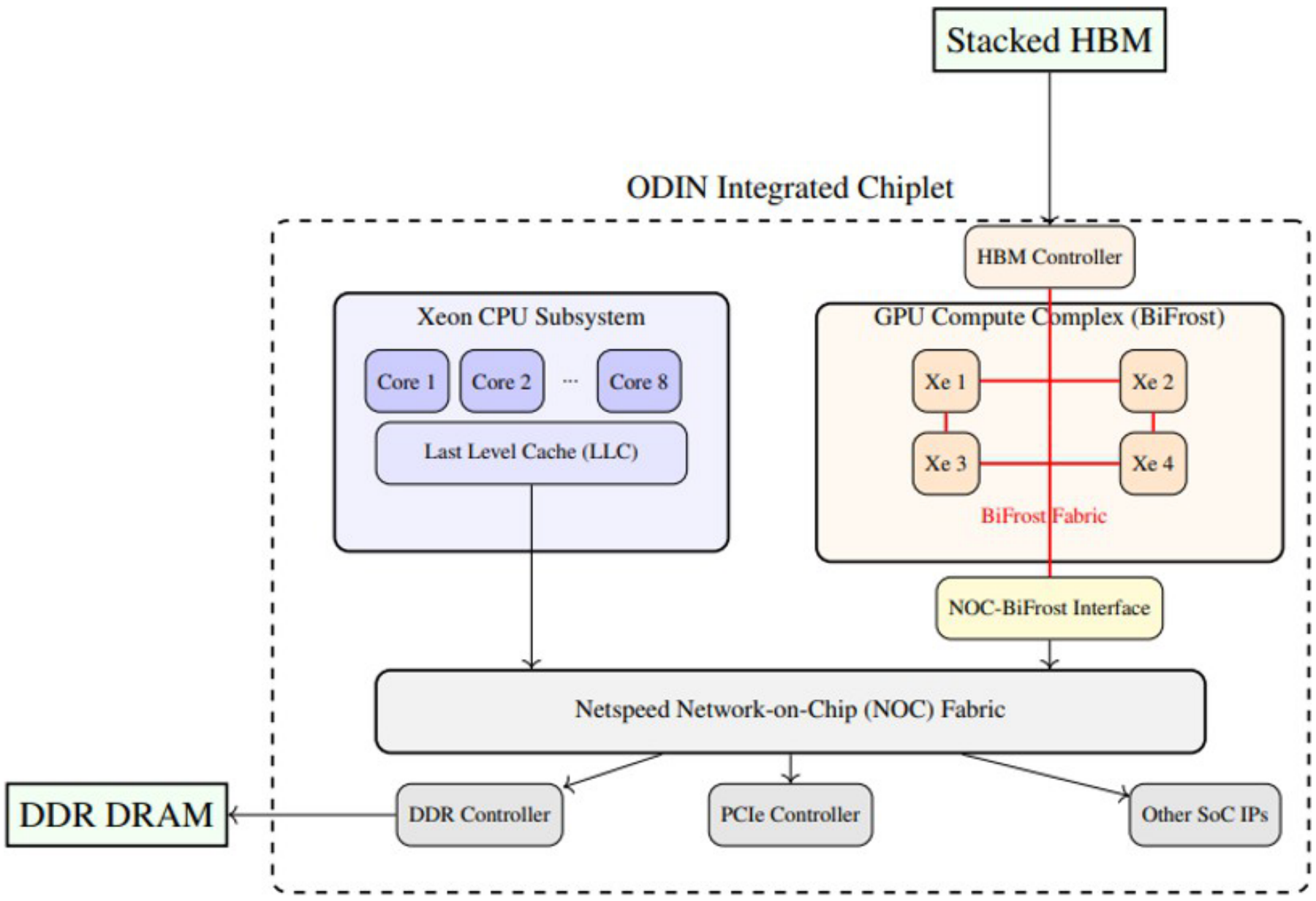

Figure 1: Odin chiplet overview.

The SoC design described as shown in Figure 2 represents a foundational building block toward the ODIN integrated chiplet architecture, rather than the final ODIN state. Individual compute and memory subsystems are composed and validated in this configuration to enable scalable integration into the broader ODIN chiplet framework. This approach allows CPU, GPU, and interconnect components to be developed and verified incrementally, while preserving architectural alignment with the eventual ODIN system composition

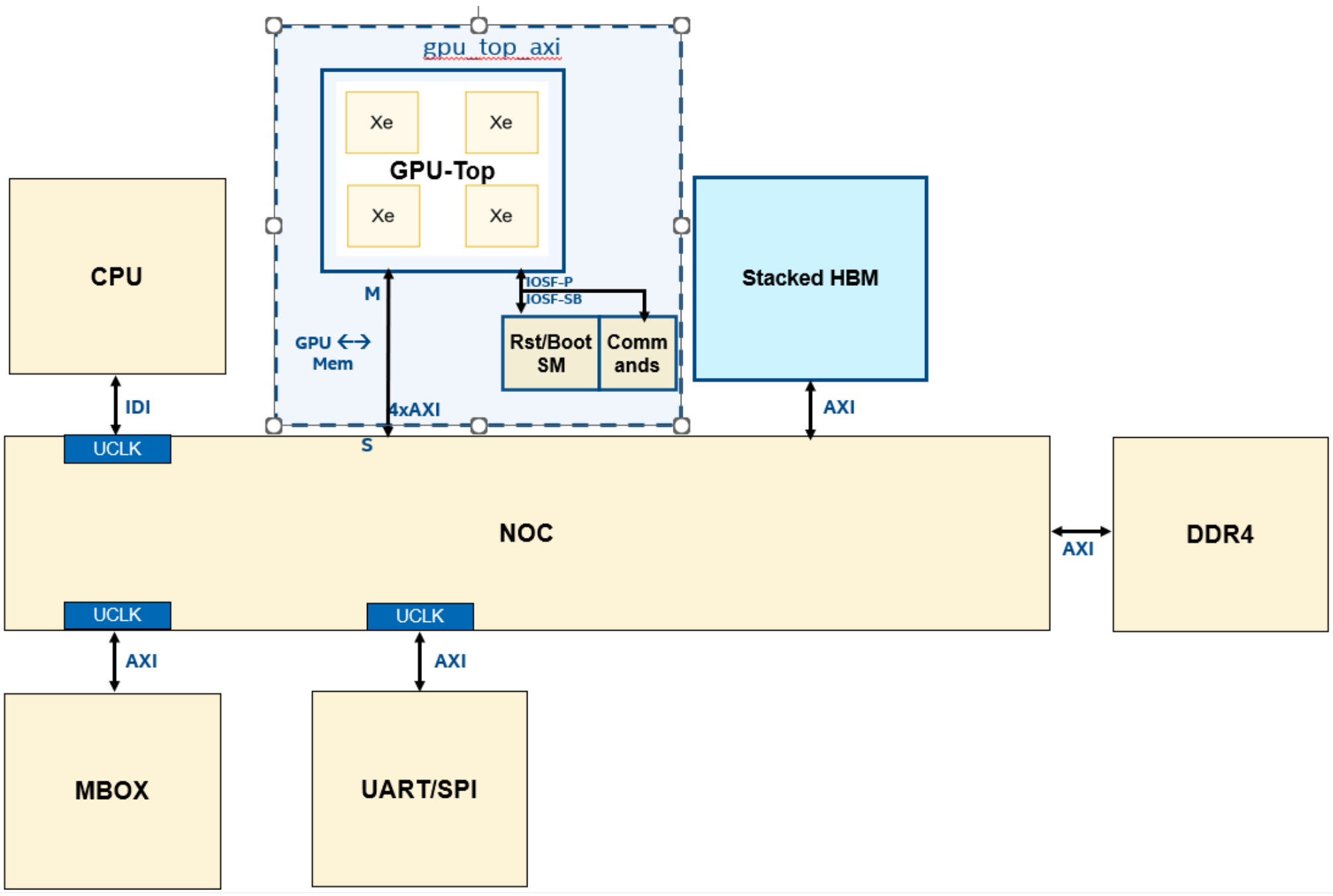

Figure 2: Top-level SoC integration showing CPU, GPU subsystem, NoC, and system memory. GPU memory and control traffic is routed through the NoC fabric to system DDR.

## 2.1 CPU Subsystem

The CPU subsystem is integrated into the NoC using coherent Intra-Die Inter-face (IDI) links. Supporting logic includes control and management block for semaphores and interrupt handling; a system controller for MMIO decode; and a power management using sideband and industry-standard Q-Channel flows. As shown in Figure 3, the subsystem connects to the NoC through coherent inter-faces.

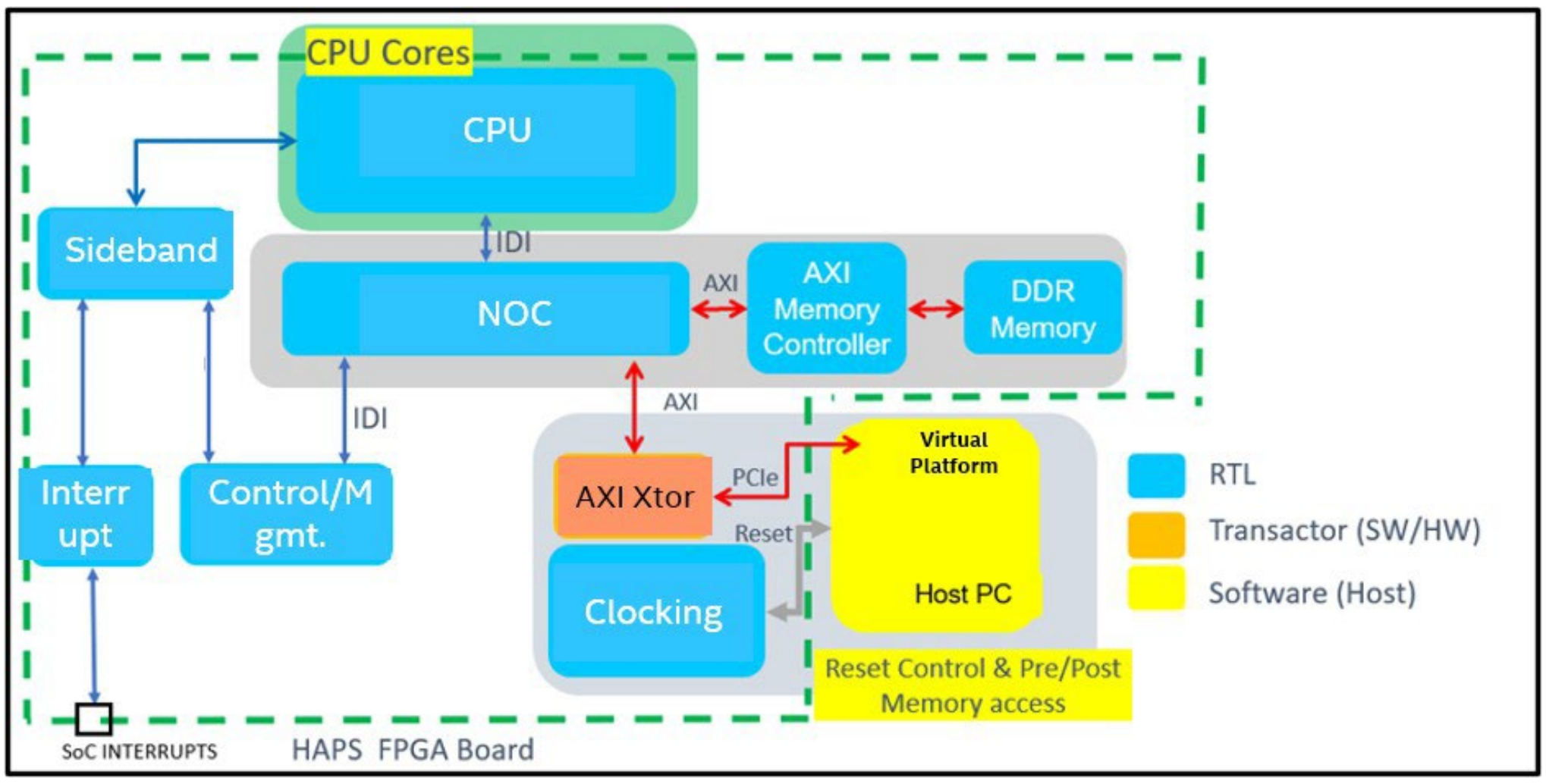

Figure 3: CPU subsystem architecture highlighting coherent IDI connectivity.

## 2.2 GPU Architecture

Conceptually, the GPU operates as a memory-input, memory-output–based processor, where workloads are expressed primarily as streams of memory transactions rather than tightly coupled control flows. This execution model enables high throughput and parallelism but places significant emphasis on the correct-ness, ordering, and performance of the memory subsystem.

The GPU IP used in this work is a synthesizable RTL design composed of multiple Xe cores, each containing several execution units (EUs). The EUs execute a mix of SIMD arithmetic operations, SIMD load/store instructions, and systolic operations, enabling efficient processing of highly parallel workloads. Work is issued to the GPU as waves of threads, with execution distributed across Xe cores and EUs to maximize utilization.

GPU execution is inherently memory-driven, with frequent accesses to sys-tem memory for instruction fetch, data reads, and result write-back. As a re-sult, the GPU relies heavily on the surrounding SoC infrastructure—including the Network-on-Chip (NoC), memory controllers, and cache hierarchy—to sustain throughput and maintain correctness. Interactions such as memory ordering, coherency, back-pressure, and response timing become critical validation points at the system level.

This tight coupling between GPU execution and system memory behavior makes GPU validation particularly sensitive to integration issues at the NoC and memory interfaces. Consequently, accurate modeling of memory traffic and deterministic reproduction of interface behavior are essential for effective pre-silicon validation. Figure 4 provides an overview of the GPU architecture and its integration context.

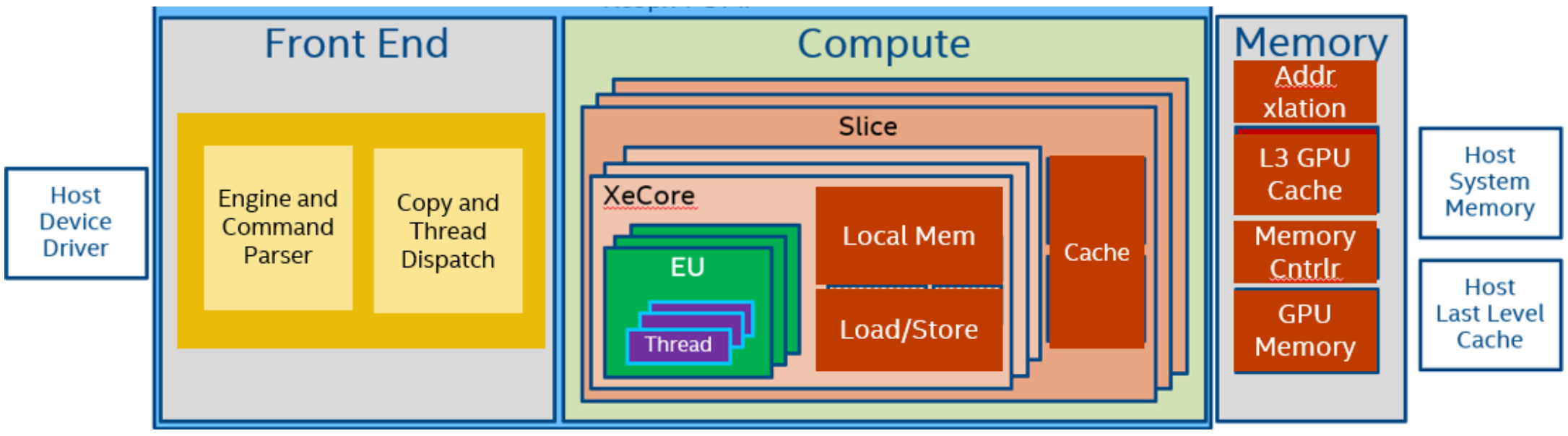

Figure 4: GPU architecture showing Xe cores, execution units (EUs), and memory hierarchy.

## 3 Validation Challenges

Several challenges emerged during CPU–GPU–NoC integration. In addition to functional complexity, the GPU execution flow relies on complex boot, power management, timing, and clocking protocols, many of which use proprietary interfaces. These protocols must be followed precisely to achieve a successful GPU bring-up and workload execution. From a full-SoC perspective, modeling and understanding these detailed sequences introduces significant overhead and complexity.

In traditional bus functional model (BFM)–based validation flows, transitioning from simulation to emulation often requires significant changes to model compilation, configuration, and execution infrastructure. These differences frequently result in maintaining separate design databases or platform-specific collateral for simulation and emulation, increasing integration overhead and making consistency across validation environments difficult to maintain. This fragmentation complicates debug correlation, slows iteration, and increases the risk of divergence between simulation and emulation results during system-level validation.

Another significant challenge during SoC-level validation is root-causing IP-level issues from full-system failures. When failures are observed at the SoC level, identifying whether the root cause lies in the IP implementation, the integration logic, or system-level interactions often requires extensive debug effort. This process typically involves back-tracking through multiple abstraction layers—from SoC-level behavior to specific IP interfaces and internal states—making debug time-consuming and resource-intensive, particularly for complex CPU–GPU interactions.

Figure 5 highlights the Replay Engine concept used to abstract GPU boot and protocol complexity at the SoC level. Additional challenges arose from randomized modeling used for metastability and pipeline staging within the GPU to improve verification coverage. While effective for coverage, this randomization complicates reproducibility and makes deterministic debug difficult. Full-chip RTL simulation is prohibitively slow for realistic workloads, while emulation lim-its internal visibility.

## 4 Replay Engine Architecture

To address these challenges, a Replay Engine was introduced to enable deterministic capture and reproduction of GPU-driven traffic across simulation and emulation. The Replay Engine interfaces at well-defined subsystem boundaries, avoiding invasive instrumentation while preserving functional fidelity.

### 4.1 Replay Engine Components

The Replay Engine captures timing-accurate waveforms at the GPU IP periphery, focusing on architecturally visible interface signals—including data, control, and response information. By preserving cycle-level behavior observed during standalone GPU IP validation, the captured trace provides a deterministic representation of the protocol interactions required for boot and workload execution.

Unlike traditional approaches that depend on a bus functional model (BFM) to consume GPU outputs and generate corresponding responses, the replay methodology does not require a live BFM during SoC replay. The captured waveform inherently contains both the request signals driven by the GPU and the corresponding responses observed at the interface boundary. During replay, these responses are re-generated by the Replay Engine in the same clock cycles in which they were originally observed, effectively emulating the presence of a responding agent without explicitly instantiating BFM collateral.

This design choice addresses multiple system-level validation challenges. First, in conventional BFM-based flows, transitioning from simulation to emulation often requires significant model and compilation changes and can result in maintaining separate simulation and emulation databases, which are difficult to keep consistent. By embedding the required stimulus and response behavior directly in the replay artifact and reusing it across platforms, the replay methodology enables a uniform validation path between simulation and emulation while avoiding duplicated integration collateral. Second, SoC-level failures are often time-consuming to root-cause back into the originating IP. Deterministic replay improves debug efficiency by enabling consistent reproduction at well-defined IP boundaries, reducing the search space when back-tracking from system behavior to specific interface regions. Finally, because replay artifacts are decoupled from dynamic BFM infrastructure, incremental design changes—such as localized IP fixes or integration updates—can be validated with minimal disruption, allowing targeted updates without reworking or re-qualifying extensive testbench collateral. In combination, replay preserves protocol correctness and determinism across simulation and emulation while reducing integration overhead and accelerating both root-cause analysis and iterative validation.

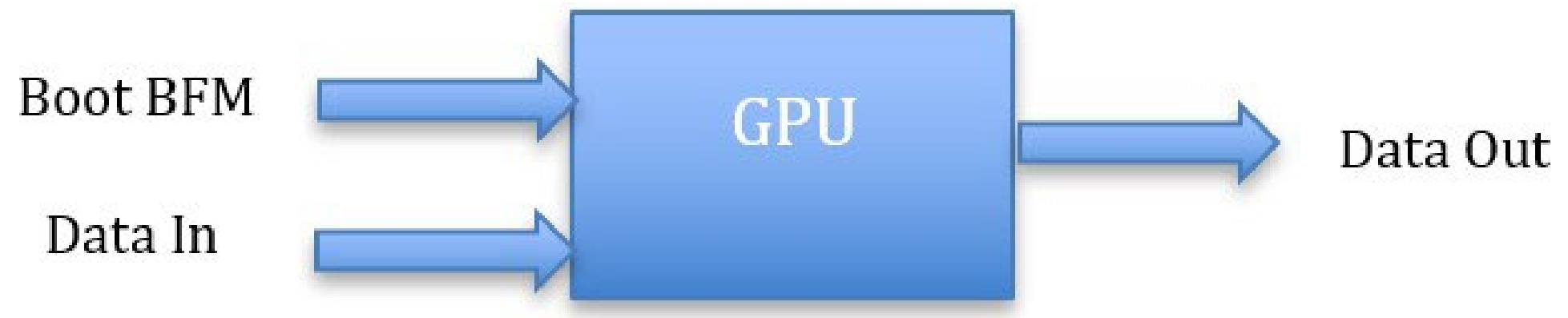

Figure 5: Replay Engine architecture capturing timing-accurate waveforms at the GPU IP periphery and converting them into ROM-initialized replay data for deterministic execution.

## 4.2 Replay Capture and ROM Initialization Flow

Replay capture is performed during standalone GPU IP validation, where the Replay Engine records waveform activity at the GPU IP periphery. The captured waveform serves as the source artifact for replay and is not consumed directly at runtime. Figure 6depicts the replay capture, conversion, and ROM initialization flow used to prepare replay data. Following capture, an offline post processing step extracts the relevant interface signals and converts the waveform activity into a cycle ordered bit representation. This conversion encodes both stimulus and corresponding response information into a compact, replay specific format suitable for storage. The resulting data represents the exact interface behavior observed during capture, preserved on a per cycle basis.

The encoded replay data is then used to initialize ROM based storage structures within the Replay Engine as part of the SoC design initialization process. During simulation or emulation, the Replay Engine reads this pre initialized ROM content to drive replay execution, eliminating any dependency on dynamic waveform files or external runtime infrastructure. By separating waveform capture, conversion, and ROM initialization from runtime replay execution, this approach enables repeatable, self contained replay across platforms while maintaining consistency with the originally captured GPU protocol behavior.

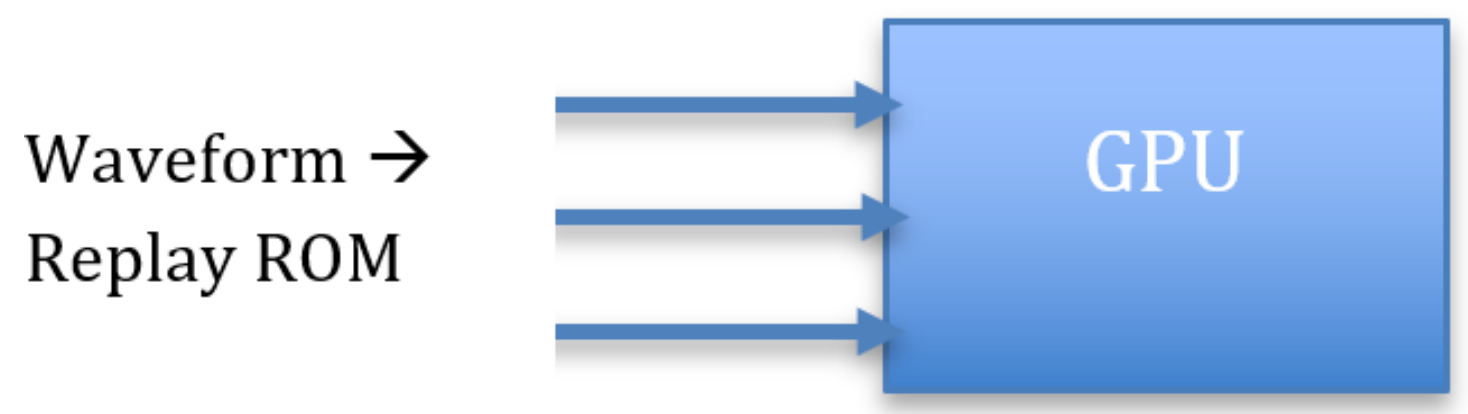

Figure 6: Replay capture and ROM initialization flow used to prepare replay data.

# 5 Simulation and Emulation Flow

## 5.1 Methodology: Replay-Driven System-Level Validation

The validation methodology employs both simulation and emulation as complementary execution platforms, built from a common design database and YAML-based configuration. Replay artifacts serve as the shared stimulus mechanism across these platforms, enabling consistent execution of GPU-driven system scenarios while allowing each environment to be used where it is most effective. Simulation is leveraged for detailed debug and waveform visibility, while emulation is used for scalable, high-speed system-level execution.

In a conventional SoC integration flow, IP subsystems typically provide testbench collateral or bus functional models (BFMs) to enable early bring-up and stimulus generation at the SoC level. Developing, validating, and maintaining this collateral often adds significant integration overhead and can delay execu-tion of the first meaningful system-level tests.

In contrast, the replay-driven methodology presented in this work eliminates the need for dedicated BFM or testbench collateral for GPU integration. The required protocol behavior already exists in the timing-accurate waveform capture performed during standalone GPU IP validation. This captured waveform, including both stimulus and corresponding responses, is converted into replay data and reused directly at the SoC level to drive boot and workload execution.

By reusing validated protocol behavior rather than re-implementing it in separate integration collateral, the methodology significantly accelerates SoC bring-up. This approach enabled the system to successfully boot and execute the first GPU-driven testcases earlier in the integration cycle, while maintaining deterministic and reproducible behavior across both simulation and emulation.

By combining detailed simulation debug with high-throughput emulation execution around a single replay mechanism, the methodology reduces debug turnaround time, avoids duplication of integration effort, and improves pre-silicon integration confidence. This replay-driven approach establishes a scalable validation framework that naturally extends to increasingly complex SoC and chiplet-based systems.

## 5.2 Simulation Flow

Simulation serves as the primary environment for functional debug and root-cause analysis, providing detailed waveform visibility across the CPU, GPU, and NoC interfaces. The Replay Engine enables deterministic execution of complex GPU protocol behavior, allowing system-level scenarios such as boot and workload execution to be analyzed with cycle-accurate signal visibility. This capability is particularly valuable for diagnosing protocol ordering, timing dependencies, and integration issues that are difficult to observe through directed testing alone. Figure 7 shows the simulation flow and output comparison against a golden reference. Figure 8

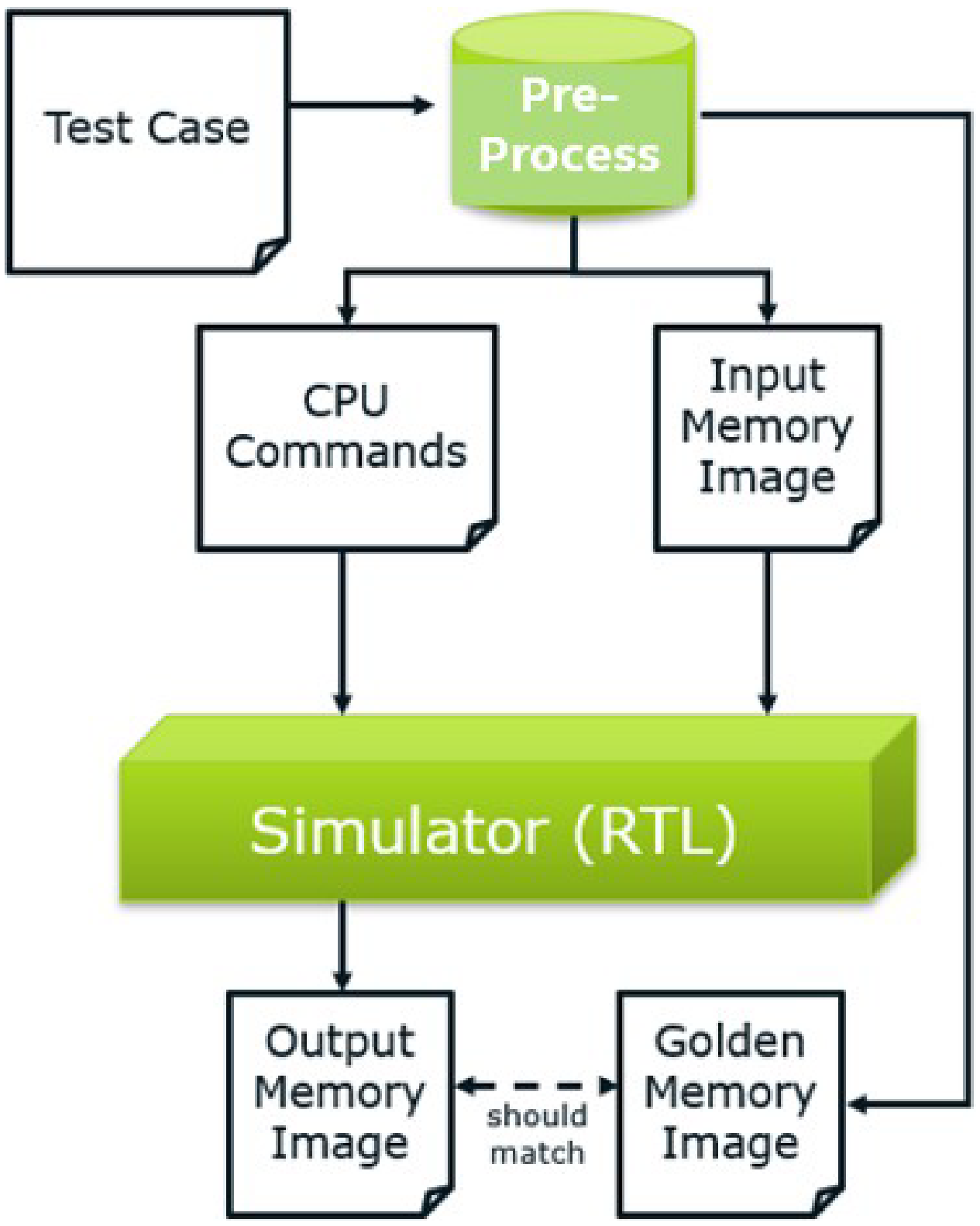

Figure 7: Simulation test flow.

| Simulation Output | Golden Output |
|---|---|
| 0000000f0000000e0000000d0000000c | 4316BC0000: 00000000010000000200000003000000 |
| 0000000b0000000a0000000900000008 | 4316BC0010: 04000000050000000600000007000000 |
| 00000007000000060000000500000004 | 4316BC0020: 08000000090000000A0000000B000000 |
| 00000003000000020000000100000000 | 4316BC0030: 0C0000000D0000000E0000000F000000 |

Figure 8: Successful simulation results using replay-driven stimulus with output comparison against a golden reference.

### 5.3 Emulation Flow

In the emulation environment, the replay artifacts generated during simulation are reused to enable high-speed execution of GPU-driven system scenarios. The Replay Engine deterministically replays captured GPU protocol behavior at the IP periphery, allowing the SoC to progress through complex sequences such as boot and workload execution without dependence on full software stacks or externally generated stimulus. This approach simplifies emulation setup while maintaining functional equivalence with scenarios validated in simulation.

Emulation complements simulation by enabling execution of long-running, system-level workloads that would be impractical to run repeatedly in RTL simulation. Although internal signal visibility is more limited than in simulation, the deterministic nature of replay allows issues observed in emulation to be reliably correlated back to simulation for detailed root-cause analysis. This combination enables rapid reproduction of integration issues at full-chip scale while preserving confidence in functional correctness.

Figure 9 shows representative emulation waveforms demonstrating successful system boot and correct memory output, illustrating how deterministic replay drives system-level execution at significantly higher performance. To demonstrate feasibility at scale, the full CPU–GPU–NoC SoC was mapped onto the emulation platform and evaluated for resource utilization. Table 1 summarizes the estimated and configured resource usage across different board configurations, showing that the design fits within available capacity while leaving margin for additional debug and infrastructure logic.

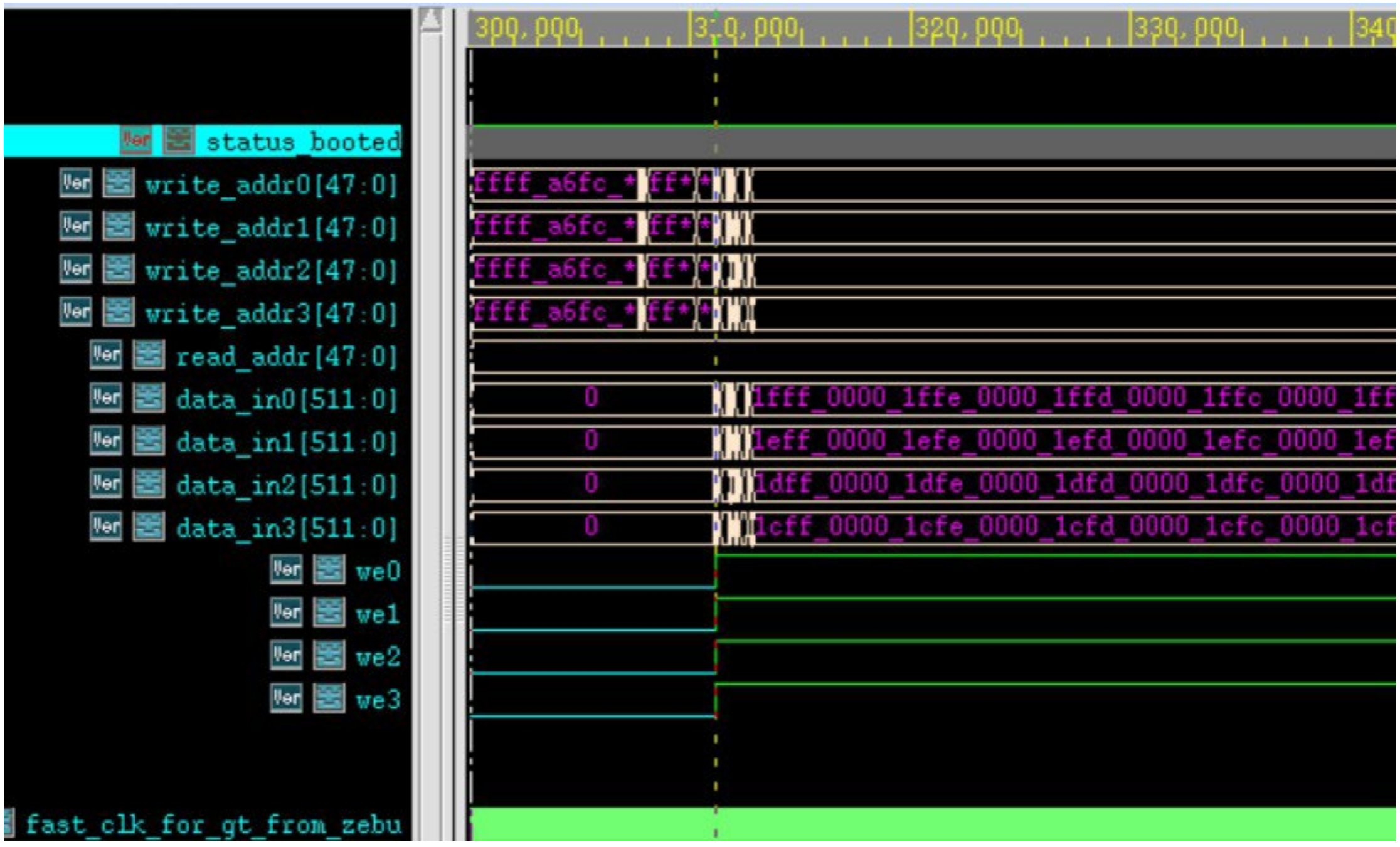

Figure 9: Emulation flow using replay artifacts to enable high-speed system-level validation.

Table 1: Estimated resource utilization for full-chip emulation on the EP1 platform

| Configuration | LUT | RAM | URAM | REG |
|---|---|---|---|---|
| Estimated Size | 87M | 25K | 503 | 36M |
| 15 Boards | 88.8% | 40.2% | 3.7% | 24.6% |
| 16 Boards (64 FPGAs) | 83.3% | 37.7% | 3.5% | 13.7% |

## 5.4 Emulation Hardware Platform (EP1)

The full-chip emulation described in this work was deployed on the EP1 emulation hardware platform, which provides the capacity and interconnect required to host the integrated CPU–GPU–NoC SoC design. EP1 serves as the execution target for the compiled design image, enabling validation of system-scale integration under realistic hardware constraints. Figure10 presents the EP1 hardware context used in this study, including the EP1 module, the target EP1 hardware platform, and the 16-board EP1 configuration shown is used to accommodate the full SoC design. Together, these views illustrate how the design scales across the emulation infrastructure to support full-chip execution while preserving consistency with the replay-driven validation flow described earlier.

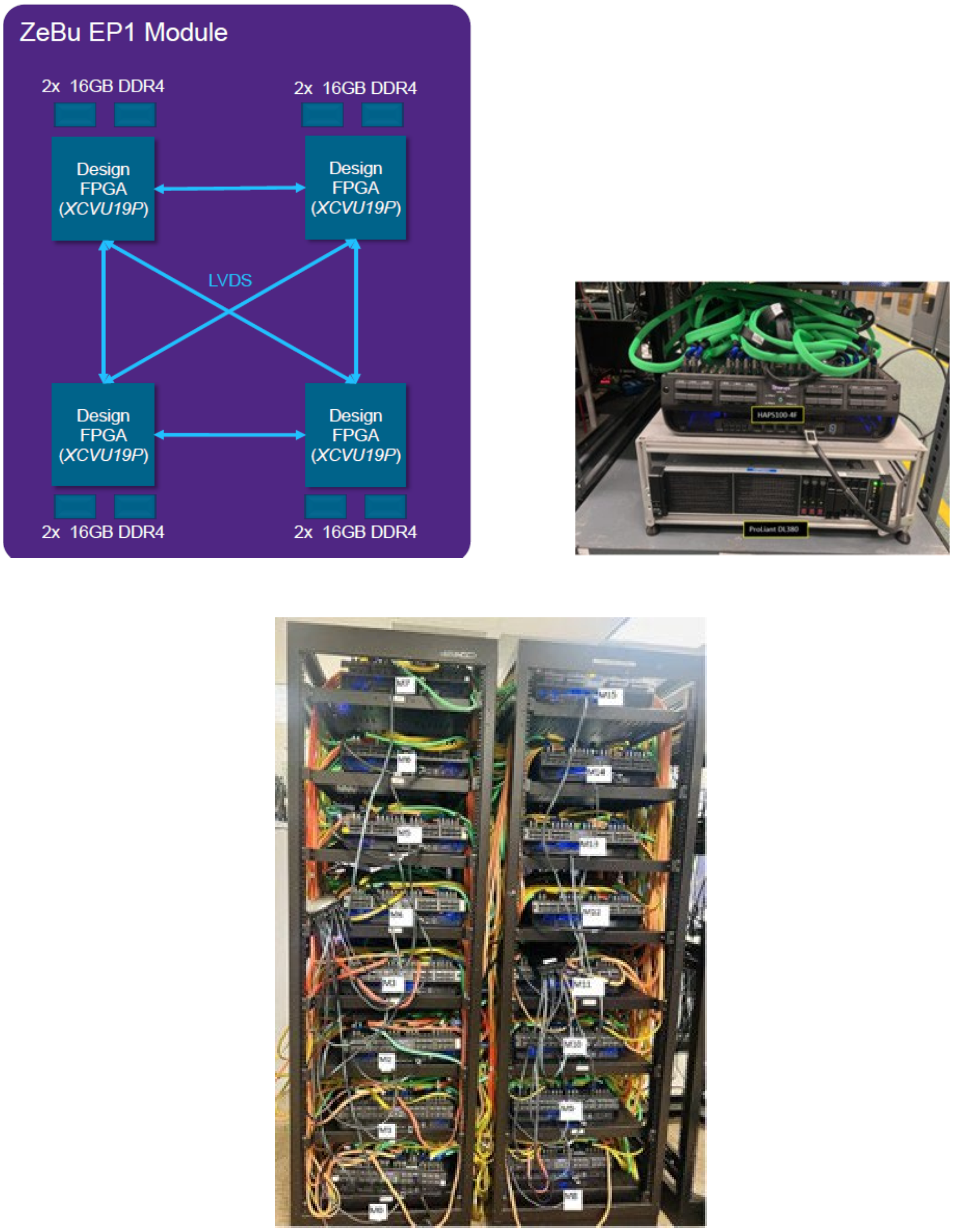

Figure 10: EP1 emulation hardware platform used for full-chip CPU–GPU–NoC system validation.

## 6 Results and Key Learnings

Using replay-driven validation, end-to-end system boot and GPU workload execution were achieved within a quarter. By reusing deterministic, protocol-accurate replay artifacts captured during standalone IP validation, the methodology eliminated the need for dedicated SoC-level testbench or BFM collateral, significantly accelerating SoC integration and enabling earlier execution of the first meaningful system-level tests.

### 6.1 Key Learnings

- Deterministic waveform capture was essential for enabling reliable replay across both simulation and emulation environments.
- Replay-based stimulus significantly reduced debug turnaround time and enabled repeatable reproduction of system-level issues.
- A reduced ROM footprint was achieved by leveraging clock-generation-based replay instead of full capture-replay of internal clocking logic.
- Automation of capture and replay logic insertion minimized manual errors and improved repeatability across validation runs.
- For emulation, clock simulation based clocking required preprocessing through the ZEMI3 flow using Synopsys BC, introducing additional setup considerations.
- The clock simulation based clocking model does not support dynamic frequency changes, requiring fixed-frequency assumptions during replay-driven emulation.
- Disabling flop randomization used for metastability validation was necessary to achieve deterministic behavior suitable for waveform replay.

## 7 Conclusion

Replay-driven validation provides a scalable, deterministic foundation for validating complex heterogeneous SoCs, where tight CPU–GPU coupling and system-level protocols make traditional validation approaches insufficient. By unifying simulation and emulation around a single, reusable replay artifact, this methodology avoids the need for separate simulation and emulation databases commonly associated with BFM-based flows, reducing integration overhead and maintaining consistency across validation platforms. In addition, deterministic replay improves debug efficiency by enabling reliable reproduction at well-defined IP boundaries, significantly reducing the effort required to root-cause SoC-level failures back to specific IP interfaces. As systems evolve toward ODIN-class chiplet architectures, replay-based validation enables repeatable, interface-accurate verification at subsystem and chiplet boundaries, establishing a practical validation framework that scales with increasing system complexity.